# Effects of deposition temperature on the mechanical and structural properties of amorphous Al-Si-O thin films prepared by RF magnetron sputtering

Stefan Karlsson[1,*], Per Eklund[2], Lars Österlund[3], Jens Birch[2], and Sharafat Ali[4*]

[1] RISE Research Institutes of Sweden AB, Department of Materials and Surface Design, Glass unit, Vejdes plats 3, SE-352 52 Växjö, Sweden

[2] Department of Physics, Chemistry and Biology, (IFM), Linköping University, SE-58183, Linköping, Sweden

[3] Dept. Materials Science and Engineering, The Ångström Laboratory, Uppsala University, P.O. Box 35, SE-75103 Uppsala, Sweden.

[4] School of Engineering, Department of Built Environment and Energy Technology, Linnæus University, SE-351 95, Växjö, Sweden

[*]Corresponding authors: stefan.karlsson@ri.se and sharafat.ali@lnu.se

## Highlights

- Increased substrate temperature gives higher hardness.
- Increasing Al/Si ratio gives a denser structure with higher hardness.
- Lower substrate temperature gives a higher crack resistance.
- Si-O-Si/Al asymmetric stretching intensity decreases linearly with Al concentration.
- Al-O-Al vibrational band shifts relate to the structural density.

## Abstract

Aluminosilicate (Al-Si-O) thin films containing up to 31 at. % Al and 23 at. % Si were prepared by reactive RF magnetron co-sputtering. Mechanical and structural properties were measured by indentation and specular reflectance infrared spectroscopy at varying Si sputtering target power and substrate temperature in the range 100 to 500 °C. It was found that an increased substrate temperature and Al/Si ratio give denser structure and consequently higher hardness (7.4 to 9.5 GPa) and reduced elastic modulus (85 to 93 GPa) while at the same time lower crack resistance (2.6 to 0.9 N). The intensity of the infrared Si-O-Si/Al asymmetric stretching vibrations shows a linear dependence with respect to Al concentration. The Al-O-Al vibrational band (at 1050 cm$^{-1}$) shifts towards higher wavenumbers with increasing Al concentration which indicates a decrease of the bond length, evidencing denser structure and higher residual stress, which is supported by the increased hardness. The same Al-O-Al vibrational band (at 1050 cm-1) shifts towards lower wavenumber with increasing substrate temperature indicating an increase in the of the average coordination number of Al.

## 1. Introduction

Aluminosilicates are composed of the 2$^{nd}$ and 3$^{rd}$ most abundant elements in the Earth's crust, Al and Si [1], and is thus one of the most important materials in geoscience [2]. It is also among the most important materials in glass technology [3]. Because of its properties, high chemical stability and refractory character, it is also of potential use as corrosion- and wear-resistant coatings [4] and gate insulator in metal oxide semiconductor (MOS) devices [5-7]. The deposition rate, composition, structure, morphology and properties of aluminosilicate thin films are influenced by their synthesis conditions [8]. Aluminosilicate thin films have previously been prepared by a variety of methods including sol-gel methods [9, 10], chemical vapor deposition (CVD) [11, 12], and physical vapor deposition (PVD) [7, 13-16], which all have their advantages and disadvantages [17]. PVD magnetron sputtering has become widely used in industry because of its versatility and proven scale-up capabilities for industrial production [18].

Amorphous oxide systems may provide useful properties for use as damage tolerant engineering materials, e.g., a-Al$_2$O$_3$ [19-21], a-SiO$_2$ [22, 23], a-Si-O-N [24], a-Mg/Ca-Si-O-N [25-27], a-Si-Al-ON [28], a-Nb$_2$O$_3$ [29-31], a-ZrO$_x$ [32], a-Ta$_2$O$_3$ [29], a-VO$_x$ [33], a-Ce$_2$O$_3$ [34] and a-TiOx [31]. Some of them might also have other useful properties, e.g., optical [14, 25, 26, 32, 35, 36], electrochemical [33, 36], electrical [37] properties thus giving multifunctional properties. Protective coatings on glass offer an interesting alternative to strengthening of glass as the strength of glass is directly determined by its inevitably many surface defects [38, 39]. The surface microcracks may be additionally deteriorated by stress-corrosion leading to additionally decrease in the glass strength [38, 40]. Therefore, surface coatings can improve mechanical, optical and chemical properties of glass products [38, 39, 41, 42]. However, the mechanical properties of amorphous Al-Si-O thin films on conventional flat glass, i.e., soda lime silicate glass, have hitherto not been studied.

The composition and structure of aluminosilicates influence its physical properties, and the primary understanding is that undoped amorphous aluminosilicate preferentially consists of corner-sharing of [SiO$_4$] and [AlO$_4$], [AlO$_5$], [AlO$_6$] units in a vitreous network [43-46]. However, there are also suggestions that 2,3 or 4-fold coordinated O may be present to fulfill the charge compensation [45, 47-49]. Löwenstein's avoidance rule suggests that the framework is constructed by Si-O-Al bridges [50] which implies that a [AlO$_4$] unit is surrounded by four [SiO$_4$] units, a concept that generally applies for aluminosilicates, e.g., zeolites. Löwenstein's avoidance rule may be violated if the Al/Si ratio is >1 [51, 52]. Vibrational spectroscopy is a useful tool to study chemical bonding in glass. The vibrational spectroscopy band assignments of silica-rich amorphous compositions are well-studied [53-56] but band assignments of Al$_2$O$_3$-rich compositions are less known [57-59].

This work is a follow-up study to our previous report on reactive radiofrequency magnetron co-sputtering synthesis of Al-Si-O coatings and their optical properties [14]. In bulk, Al$_2$O$_3$-SiO$_2$, have shown to have good crack resistance [60] which makes the investigation of Al-Si-O thin films interesting for damage resistance, in addition to wear and corrosion resistance. In this study, we analyze the mechanical properties and the structure of Al-Si-O thin films by indentation and infrared spectroscopy. Furthermore, we explore the possibility of decreasing the coating process temperature as a route towards energy efficient magnetron sputtering by investigating the deposition temperatures 500, 300 and 100 °C.

## 2. Experimentals

### 2.1 Materials and Deposition

The substrate was a 1 mm float glass substrate in the form of conventional microscope slides. The typical composition (given in wt%) is 72.3% SiO$_2$, 0.5% Al$_2$O$_3$, <0.02% Fe$_2$O$_3$, 13.3% Na$_2$O, 8.8% CaO, 0.4% K$_2$O and 4.3% MgO, i.e., a soda lime silicate glass.

Al-Si-O thin films were deposited on the atmospheric side of float glass by radio frequency (RF) magnetron reactive co-sputtering using elemental targets of Al and Si in ultra-high vacuum (UHV). The deposition system is described in detail elsewhere [61]. An oxygen and argon mixture were kept constant with a total gas flow of 40 mL/min ($O_2$ mL/min and Ar mL/min) at standard temperature and pressure, STP, conditions. The Al target power was kept constant at 100 W and the Si target power was varied between 40, 60 and 80 W. In addition, the substrate temperature was varied between 100, 300 and 500 °C giving a sample matrix of 9 samples, see Table 1. Complete details of the physical vapor deposition processing are given in ref. [14].

## 2.2 Nanoindentation

Nanoindentation with a Berkovich tip was used for measuring hardness ($H$) and reduced elastic modulus ($E_r$) according to the Oliver and Pharr method [62], where $H$ is defined by $H = \frac{F_m}{A_p}$, with $F_m$ is the maximum applied load and $A_p$ the projected contact area. $A_p$ is calculated by polynomial fitting: $A_p(h_c) = C_0 h_c^2 + C_1 h_c^1 + C_2 h_c^{1/2} + C_3 h_c^{1/4} + \cdots + C_8 h_c^{1/128}$, where $C_x$ is the indenter specific factors with $C_0$ being 24.56 for a perfect Berkovich tip and $h_c$ is the real contact depth that compensates for the sink-in effect by $h_c = h_m - \varepsilon \frac{F_m}{S}$, where $h_m$ is the maximum penetration depth, $\varepsilon$ is a tip factor for a Berkovich tip ($\varepsilon = 0.72$) and $S$ is the stiffness defined by the slope upon unloading, $S = \frac{\partial P}{\partial h}$. The reduced elastic modulus is defined as $E_r = \frac{\sqrt{\pi}}{2\beta} \frac{S}{\sqrt{A_p}}$ where $\beta$ is a geometrical tip factor, which is $\beta = 1.034$ for a Berkovich tip.

The samples were cleaned with ethanol and laboratory tissue before measurements. The nanoindenter instrument was an Anton Paar NHT[2] and the measurements were run with the loads 1, 5, 10, 15, 25, 50 and 75 mN. 20 indents were made for each load. In some cases, one indentation outlier datapoint, or for the 1 mN load, sometimes two datapoints, was removed from the analysed data due to unrealistic scattering. The measurement settings were the following: 10 Hz acquisition rate, loading/unloading rate two times the load per minute, 10 s holding time at max load, 0.2 μm/N frame compliance, and 500 μN/μm stiffness threshold. The Berkovich tip geometry was calibrated using a standard reference material of fused silica and the measurements were performed at ambient pressure at a temperature of 22±2 °C and a relative humidity of 30 ± 15%.

## 2.3 Microindentation and Crack Resistance

The crack resistance was determined from microindentation measurements using a Vickers indenter tip that were performed using a Micro-Combi Tester from CSM Instruments. 15 indents for each load were made using a Vickers diamond tip. The microindentations were run with the following settings: 10 Hz acquisition rate, loading/unloading rate two times the load per minute, 15 s holding time, 8 μm/min approach speed, 16.6 μm/min retract speed, 30 mN contact force, and 25 mN/μm contact stiffness threshold. The Vickers indenter tip was calibrated using a standard steel reference material.

The crack resistance method is described in detail elsewhere [20], and follows the original procedures [63, 64]. From the microindentation imprints the probability of radial crack initiation ($PCI$) was calculated from counting the radial corner cracks and the results were fitted to the Weibull sigmoidal function,

$$PCI = 1 - exp^{-\left(\frac{x}{x_c}\right)^m} \tag{1}$$

where x is the load, $x_c$ the characteristic values and m is the Weibull modulus. The crack resistance, $CR$, is then defined as the load when the $PCI$ is equal to 0.5 (50% probability). All microindentations were performed in an environment with a temperature of 22±1 °C and a relative humidity of 20 ± 10%.

The error of the Weibull fit was calculated by the corresponding load deviation as given by the root-mean-square deviation, *RMSD*, of the *PCI*,

$$RMSD = \sqrt{\frac{\sum_i^N (PCI_{real} - PCI_{fit})^2}{N}} \qquad (2)$$

where $PCI_{real}$ is the experimentally determined PCI from the crack resistance test, $PCI_{fit}$ is the fitted *PCI*, *N* is the number of different loads tested in the series and *i* is the specific load tested in the series. The error of CR is estimated to be in the range 10-20% [65].

### 2.4 Infrared Spectroscopy

Specular Reflectance FT Infrared Spectroscopy was measured in the range from 600 to 4000 cm⁻¹ using a Bruker IFS 66v/S employing a liquid $N_2$ cooled MCT detector. The angle of incidence was 30 degrees. Spectra were averaged over 33 reflectance scans at a resolution of 2 cm⁻¹. A microscope slide was used as background. The reflectance spectra were cut to the region of interest, 600-1400 cm⁻¹, and were then baseline-corrected with an adaptive baseline (coarseness 75 and offset 0) which was individually applied on all spectra using Spectragryph [66]. Reflectance spectra were converted to Kubelka-Munk (KM) function by the equation $K/s = \frac{(1-R)^2}{2R}$ [67]. OriginPro version 10 was used for deconvolution of the baseline-corrected KM transformed spectra.

## 3. Results

### 3.1 Hardness and Reduced Elastic Modulus

For measuring hardness (*H*) and reduced elastic modulus (*E*ᵣ) of coatings [68] as rule of thumb, the penetration depth should not exceed >10% of the coating thickness to avoid substrate effects [69]. *H* and *E*ᵣ was measured using nanoindentation methodology, which is fairly stable down to loads of the order 1 mN for hard materials such as amorphous oxides. Loads up to 75 mN were measured to explore whether the results are consistent also at higher loads. The results are presented in Fig. S1 and S2

*H* increases with the substrate temperature as a function of the Al concentration, see Fig. 1a and Tab. 1. Fig.S3a also shows *H* at different substrate temperatures. *E*ᵣ on the other hand increases linearly for Si target powers of 80 and 60 W as a function of Al concentration, and at 40 W *E*ᵣ has a maximum for 300 °C. In Fig. S3b *E*ᵣ is also presented at different substrate temperatures, where it follows an increasing trend with the Al concentration. *H* is less easily described in series of the substrate temperatures where it is less obvious with a trend as a function of the Al-concentration, see Fig. S3.

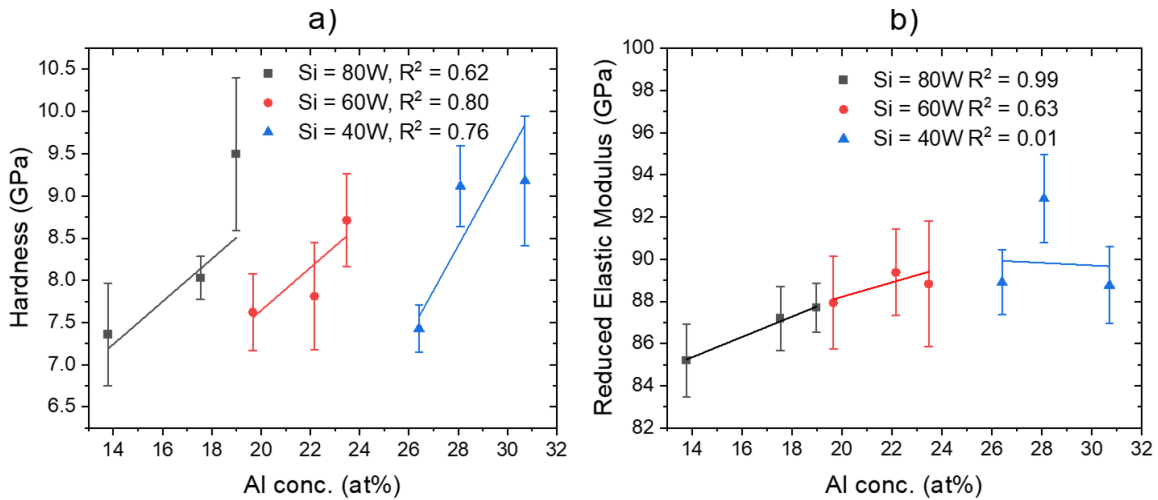

Figure 1: a) $H$ and b) $E_r$ in Si target Power series as a function of the Al concentration in at% (as taken from [14]).

Table 1 presents the parameters $H/E_r$, $U_E/U_T$ (elastic energy of recovery) and $H^3/E_r^2$ used for assessing the tribological properties where, as a rule of thumb, materials having $H/E_r$>0.1 are wear resistant [39, 70]. $H/E_r$ can then also be directly connected to the elastic recovery energy ($U_E/U_T$) in the indentation process given by

$$\frac{U_E}{U_T} = \kappa \frac{H}{E_r} \qquad (2)$$

where $\kappa$ is a proportionality factor, $\kappa \approx 5.17$ in the range 0.08-0.12 for $H/E_r$ [71]. $H^3/E_r^2$ is a measure of the resistance to plastic deformation [65, 72].

The synthesized Al-Si-O thin films exhibit good tribological properties and for those prepared at a substrate temperature of 500 °C exhibit excellent tribological properties, i.e., $H/E_r \approx 0.1$ [39]. The trends of $H/E_r$, $U_E/U_T$, and, $H^3/E_r^2$ follow the same as $H$ as a function of Al concentration, and thus increases with increasing substrate temperature. $H/E_r$ and $H^3/E_r^2$ can in general be seen as a proxy for the fracture toughness of hard coatings [73], but it is unclear how the ductility affects the proxy.

Table 1: Target power (P), substrate temperature, chemical composition, thickness ($d$) (as taken from [14]) as well as nanoindentation mechanical properties: hardness ($H$), reduced elastic modulus ($E_r$), $H/E_r$, $U_E/U_T$ and $H^3/E_r^2$ of Al-Si-O thin films.

| Si P (W) | Al P (W) | T (°C) | Al (at%) | Si (at%) | O (at%) | Al/Si ratio | $d$ (nm) | $H$ (GPa) | $E_r$ (GPa) | $H/E_r$ | $U_E/U_T$ (%) | $H^3/E_r^2$ (GPa) |
|---|---|---|---|---|---|---|---|---|---|---|---|---|
| 80 | 100 | 100 | 13.78 | 22.67 | 63.54 | 0.61 | 392.5 ± 0.020 | 7.36 ± 0.61 | 85.20 ± 1.72 | 0.086 | 44.7 | 0.055 |
| 60 | 100 | 100 | 19.66 | 18.79 | 61.54 | 1.05 | 286.28 ± 0.016 | 7.62 ± 0.45 | 87.94 ± 2.19 | 0.087 | 44.8 | 0.057 |
| 40 | 100 | 100 | 26.41 | 12.95 | 60.64 | 2.04 | 203.12 ± 0.012 | 7.43 ± 0.28 | 88.91 ± 1.55 | 0.084 | 43.2 | 0.052 |
| 80 | 100 | 300 | 17.54 | 21.07 | 61.39 | 0.83 | 317.47 ± 0.015 | 8.03 ± 0.25 | 87.18 ± 1.51 | 0.092 | 47.6 | 0.068 |
| 60 | 100 | 300 | 22.16 | 17.50 | 60.34 | 1.27 | 241.42 ± 0.012 | 7.81 ± 0.64 | 89.38 ± 2.06 | 0.087 | 45.2 | 0.060 |
| 40 | 100 | 300 | 28.09 | 12.64 | 59.27 | 2.22 | 165.41 ± 0.009 | 9.12 ± 0.48 | 92.88 ± 2.09 | 0.098 | 50.7 | 0.088 |
| 80 | 100 | 500 | 18.98 | 20.78 | 60.23 | 0.91 | 281.11 ± 0.014 | 9.49 ± 0.90 | 87.70 ± 1.17 | 0.108 | 56.0 | 0.111 |
| 60 | 100 | 500 | 23.47 | 16.28 | 60.28 | 1.44 | 208.42 ± 0.009 | 8.71 ± 0.55 | 88.83 ± 2.97 | 0.098 | 50.7 | 0.084 |
| 40 | 100 | 500 | 30.71 | 10.17 | 59.12 | 3.02 | 144.79 ± 0.016 | 9.18 ± 0.77 | 88.77 ± 1.82 | 0.103 | 53.5 | 0.098 |

### 3.2 Crack Resistance

The Crack resistance ($CR$) as a function of Al concentration is shown in Fig. 2 and the probability of crack initiation ($PCI$) as a function of load is shown in Fig. S4. In contrast to $H$, $E_r$ and the tribological parameters, $CR$ decreases with increasing Al concentration. $CR$ can be seen as a measure of the ductility, thus its ability to protect the coated glass from hard contact damage. Ductility covers both densification and plastic deformation (the latter is given by $\frac{U_P}{U_T} = 100 - \frac{U_E}{U_T}$, see Table 1). Since the trend is opposite to the tribological parameters the *CR* is in this case governed by densification and thus a low Al/Si ratio facilitates a more open structure that can densify. The Al-Si-O coated glass substrates frequently showed delayed cracking, which is a relatively common feature for bulk glasses [74]. We recorded both the immediate *CR*, and the delayed *CR* which reduced the *CR* by about 5 to 10%, see Table 2.

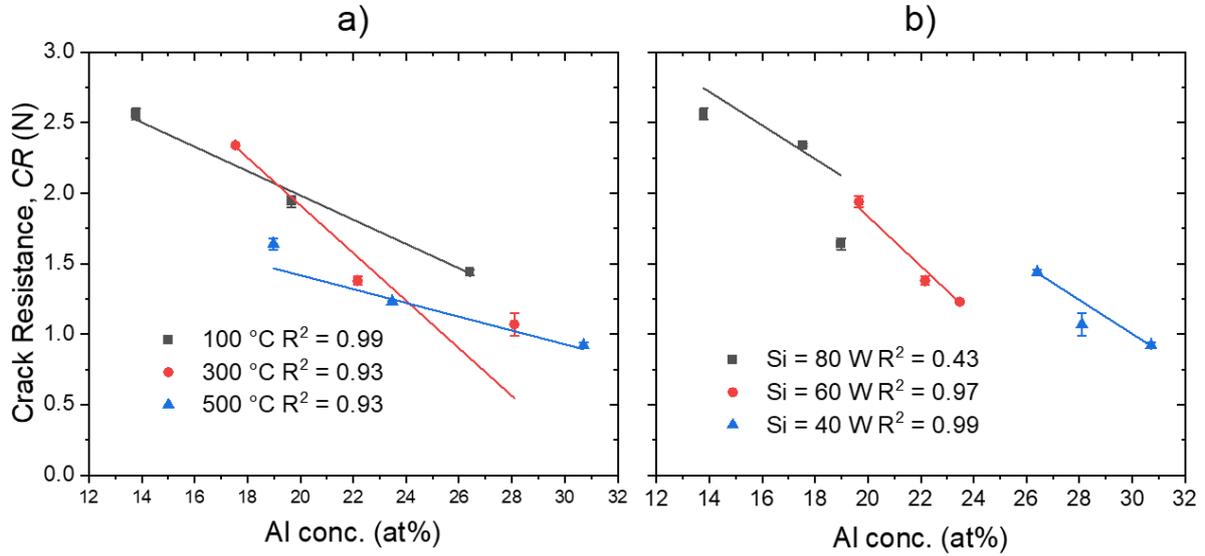

Figure 2: Crack resistance (CR) as a function of Al concentration in a) series of substrate temperature and b) in series of Si target power.

Table 2: Direct and delayed crack resistance (*CR*) including Weibull fitting parameters, the characteristic crack resistance ($x_c$) and Weibull modulus (*m*) (q.v., Eq. 1). The error represents the Weibull fit error.

| | | **Direct Crack resistance** | | | | **Delayed Crack resistance** | | | |
|---|---|---|---|---|---|---|---|---|---|
| **T (° C)** | **Si TP (W)** | *CR* (N) | Error (N) | $x_c$ (N) | *m* | *CR* (N) | Error (N) | $x_c$ (N) | *m* |
| **100** | 80 | 2.56 | ± 0.04 | 2.77 | 4.54 | 2.45 | ± 0.02 | 2.65 | 4.72 |
| **100** | 60 | 1.94 | ± 0.04 | 2.18 | 3.22 | 1.91 | ± 0.01 | 2.08 | 4.22 |
| **100** | 40 | 1.44 | ± 0.02 | 1.60 | 3.45 | 1.32 | ± 0.02 | 1.46 | 3.56 |
| **300** | 80 | 2.34 | ± 0.01 | 2.51 | 5.15 | 2.07 | ± 0.02 | 2.24 | 4.61 |
| **300** | 60 | 1.38 | ± 0.03 | 1.54 | 3.35 | 1.25 | ± 0.02 | 1.36 | 4.25 |
| **300** | 40 | 1.07 | ± 0.08 | 1.21 | 3.03 | 0.96 | ± 0.01 | 0.99 | 12.29 |
| **500** | 80 | 1.64 | ± 0.04 | 1.77 | 4.93 | 1.53 | ± 0.04 | 1.65 | 4.76 |
| **500** | 60 | 1.23 | ± 0.01 | 1.32 | 5.06 | 1.20 | ± 0.01 | 1.28 | 5.73 |
| **500** | 40 | 0.92 | ± 0.02 | 0.99 | 4.72 | 0.81 | ± 0.01 | 0.86 | 6.18 |

### 3.3 FT-IR Spectroscopy

Fig. 3 shows specular FTIR spectra for thin films prepared at different substrates temperatures. Raw FTIR spectra are shown in Fig. S5. The trends in Fig. 3 are very clear at all substrate temperatures. The broad peak centered around 1000 $cm^{-1}$ consists of at least two different vibrational modes, asymmetric stretching vibrations of Si-O-Si/Al [51, 75-77] at 1000 $cm^{-1}$ and Al-O-Al [57, 58] at 1050 $cm^{-1}$. The intensity of the 1000 $cm^{-1}$ band increases drastically with increasing Si target power, and the relative FWHM/Height ratio decreases with increasing substrate temperature, i.e., the peak gets more narrow, both trends suggesting increased structural order on a local molecular scale, even though the thin films are X-ray amorphous [14]. The smaller peak at 800 $cm^{-1}$ is assigned to Si-O-Si bending and/or stretching vibrations [75] and increases with increasing Si target power. The band at 1200 $cm^{-1}$ which is assigned to asymmetric Al-O-Al stretching modes, increases with increasing Si target power and generally red-shifts with increasing power [57]. At 80 W Si target power, a band at 1350 $cm^{-1}$ consistently appears at all substrate temperatures, and is tentatively assigned to O-Al=O- chains [59]. Based on these band assignments we have chosen to deconvolute the wavenumber range 600-1400 $cm^{-1}$, see Fig. S6-S8, using the band assignments given in Table 3. The deconvoluted data is used and discussed in relation to the mechanical properties in section **4**.

Table 3: Summary of the band assignments in the infrared range 600-1400 $cm^{-1}$.

| Band | Assignment | References | Deconvolution center constraints |
|---|---|---|---|
| 750 $cm^{-1}$ | Si-O-Si bending and/or stretching | [75] | Fixed at ca. 750 $cm^{-1}$ |
| 1000 $cm^{-1}$ | Si-O-Si/Al asymmetric stretching | [51, 76, 77] | Bound to range 940-1040 $cm^{-1}$ |
| 1050 $cm^{-1}$ | Al-O-Al stretching | [57, 58] | Bound to range 1030-1080 $cm^{-1}$ |
| 1200 $cm^{-1}$ | Al-O-Al asymmetric stretching | [57] | - |
| 1350 $cm^{-1}$ | O-Al=O- stretching | [59] | Fixed at ca. 1350 $cm^{-1}$ |

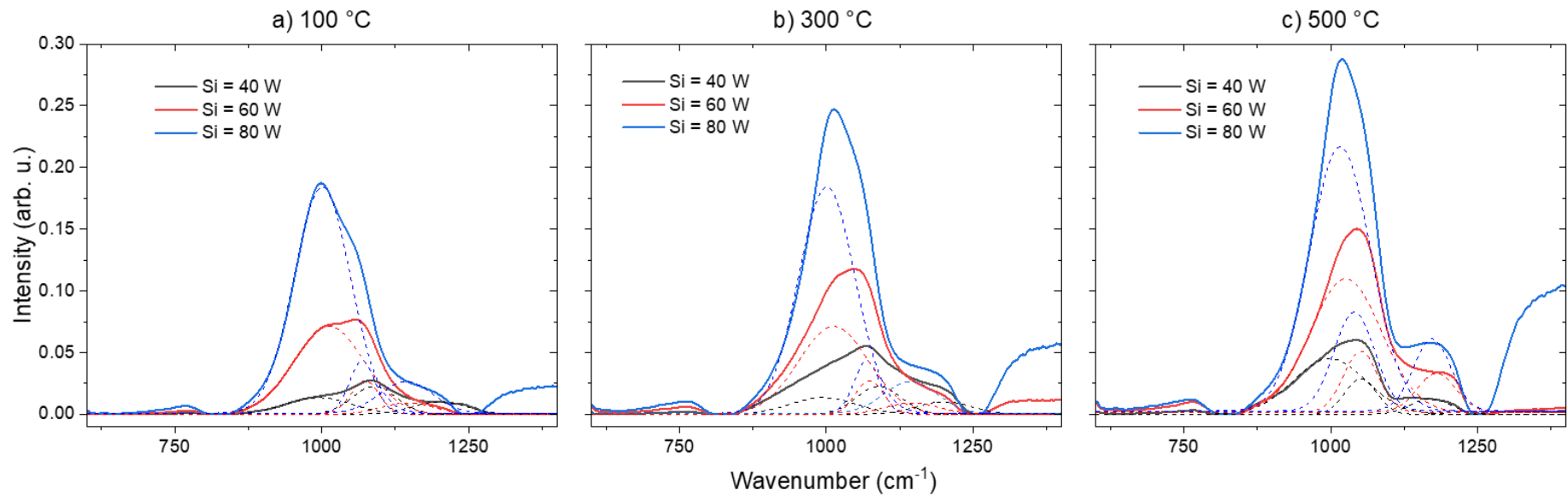

Figure 3: FTIR spectroscopy spectra of reactive co-sputtered Al-Si-O thin films prepared with substrate temperatures a) 100 °C, b) 300 °C and c) 500 °C. Solid lines are spectra after baseline-correction and KM-transformation, see section **2.4**, and dashed lines are deconvolutions of the 1000, 1050 and 1200 cm$^{-1}$ bands.

## 4. Discussion

The composition-property trends are relatively clear; however, we also note that some of the trends are nonlinear, as can be seen in Fig. 1 and 2. One possible explanation could be that it is a mix of two network former effect. Another possibility is that Si preferentially forms tetrahedral units [78, 79]. Al on the other hand is known to exist in 4, 5 or 6 coordination alternatives and it is reasonable to assume that the average coordination increases with increasing Al concentration [3, 80, 81]. The increase of the 1350 cm⁻¹ band indicates increase of -O-Al=O- chains at higher substrate temperature and higher Si target power indicating a possible increase of the local order and/or degree of crystallinity. From the measured composition given in Table 1 we can calculate that the mismatch to stoichiometry, i.e., missing O atoms is in the range from 2.5% to 10%. The non-stoichiometry increases with substrate temperature but there is no clear trend with the Si target power.

The compositions and IR intensity trends are clear, especially the 1000 cm⁻¹ band assigned to the Si-O-Si/Al asymmetric stretching. The deconvoluted intensity data for this band has a quite linear correlation with the Al concentration, see Fig. 4a. Furthermore, the deconvoluted 1050 cm⁻¹ Al-O-Al band shift relatively linearly correlated to the Al concentration giving a blue-shift of the peak with increasing Al concentration, see Fig. 4b. Therefore, the Si concentration as well as $H$, $E_r$ and $CR$ correlates linearly with both the 1000 cm⁻¹ intensity and 1050 cm⁻¹ band shift. The blue shift can be observed as the Al concentration increases in the respective series (Fig. 4) indicating a decrease of the average bond length [53], thus the thin films become denser, as is manifested by the increase of $H$ (Fig. 2a). On the other hand, the red-shift of the 1050 cm⁻¹ band that can be observed with increasing substrate temperature (see Fig. 4), can be explained by an increasing average coordination number of Al that counteracts the supposed blue-shift as the PVD process temperature increases the residual stress and thus also the density of the thin film [82], which we can see in the increase of the $H$ and $E_r$. An increase in the Al average coordination number would in fact decrease also the $CR$ as it becomes a less adaptive network [83]. In conclusion, we hypothesize than an Al-Si-O thin film structure develops at low Al/Si ratio that follows the Löwenstein's avoidance rule comprising Al-O-Si bridges [51, 52]. However, with increasing Al/Si ratio Löwenstein's avoidance rule is violated and AlOₓ-rich clusters are formed that promote crystallization and increase the Al average coordination number.

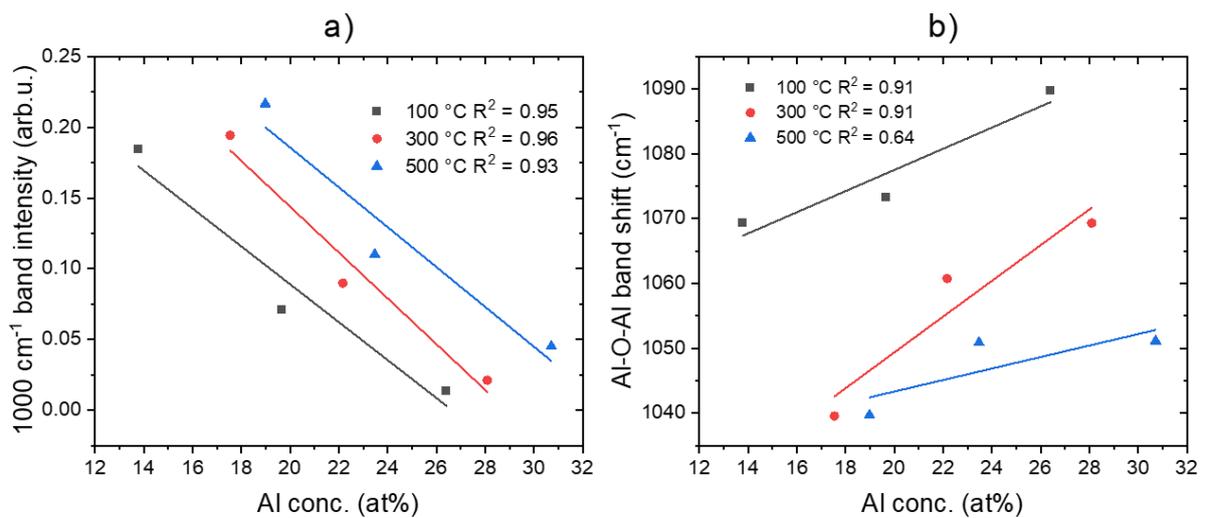

Figure 4: a) Deconvoluted 1000 cm⁻¹ band intensity and b) deconvoluted 1050 cm⁻¹ band shift as a function of Al concentration. See deconvoluted spectra in Fig. S6-S8.

$Al_2O_3$-$SiO_2$ in bulk has shown increasing trends of $H$ and $E$ with increasing Al concentration [60], consistent with our results. Bulk $CR$ has also been shown to increase with the $Al_2O_3$ concentration up to 60 mol%, in contrast to our work where the $CR$ is reduced with increasing Al concentration. This can be explained by the variable density of our thin films in combination with lower thin film thickness with increasing the Al concentration, see Table 1 and ref [14]. $CR$ of bulk and Al-Si-O thin films can thus not be compared as it is in the latter case to a large extent governed by the substrate beneath, in our case a soda-lime-silicate with $CR \approx 0.7$ [20, 84].

## 5. Conclusions

Al-Si-O amorphous thin films were synthesized using RF magnetron sputtering with different Si target power and substrate temperatures giving a 3×3 matrix. The thin films having Al/Si ratios of approximately 0.6 to 3 exhibits trends in both their mechanical properties as measured by indentation and in their IR spectra. From this study we can draw the following conclusions:

- Higher substrate temperature and lowering the Si target power gives a higher Al/Si ratio and a denser structure that results in higher hardness and reduced elastic modulus but less crack resistance.
- A strong vibrational band assigned to the Si-O-Si/Al asymmetric stretching gives high intensity to composition correlation and thus also to the properties.
- A vibrational band assigned to the Al-O-Al shift shows around 1050 $cm^{-1}$ blue-shifts with increasing Al concentration indicating a decrease in the average bond length, manifesting a denser structure that is evidenced by an increase of the hardness.
- The Al-O-Al band around 1050 $cm^{-1}$ red-shifts with increasing substrate temperature, indicating an increase of the Al average coordination number.

The IR band assignments for high Al/Si ratios are not well established in literature, therefore, further studies are needed to complete the understanding of the Al-O vibrational modes.

## 6. Acknowledgements

SK and PE acknowledge financial support by the Swedish Energy Agency (Grant No. 52740-1). SK acknowledge financial support by FORMAS, the Swedish Research Council for Sustainable Development (Grant No. 2018-00707). SA acknowledge the financial support from the KKL Advanced Materials, LNU (Grant No. 87202002) and Crafoord Foundation (Grant No. 2022-0692).

## 7. Declaration of competing interest

PE, JB, and SA are co-founders and co-owners of a startup company aiming to commercialize hard transparent coatings. The other authors declare that they have no known competing financial interests or personal relationships that could have appeared to influence the work reported in this paper.

## 8. CRediT author contributions

SK: Methodology, Validation, Formal analysis, Investigation, Resources, Data Curation, Writing – Original Draft, Writing – Review & Editing, Visualization. PE: Conceptualization, Methodology, Resources, Writing – Review & Editing. LÖ: Methodology, Resources, Writing – Review & Editing. JB: Conceptualization, Methodology, Writing – Review & Editing, SA: Conceptualization, Methodology, Investigation, Resources, Writing – Review & Editing.

## 9. Data availability

The data that support the findings of this study are available from the corresponding authors upon reasonable request.

# Supplementary Materials

# Effects of deposition temperature on the mechanical and structural properties of amorphous Al-Si-O thin films prepared by RF magnetron sputtering

Stefan Karlsson[1,*], Per Eklund[2], Lars Österlund[3], Jens Birch[2], and Sharafat Ali[4*]

[1] RISE Research Institutes of Sweden AB, Department of Materials and Surface Design, Glass unit, Vejdes plats 3, SE-352 52 Växjö, Sweden

[2] Department of Physics, Chemistry and Biology, (IFM), Linköping University, SE-58183, Linköping, Sweden

[3] Dept. Materials Science and Engineering, The Ångström Laboratory, Uppsala University, P.O. Box 35, SE-75103 Uppsala, Sweden.

[4] School of Engineering, Department of Built Environment and Energy Technology, Linnæus University, SE-351 95, Växjö, Sweden

[*]Corresponding authors: stefan.karlsson@ri.se and sharafat.ali@lnu.se

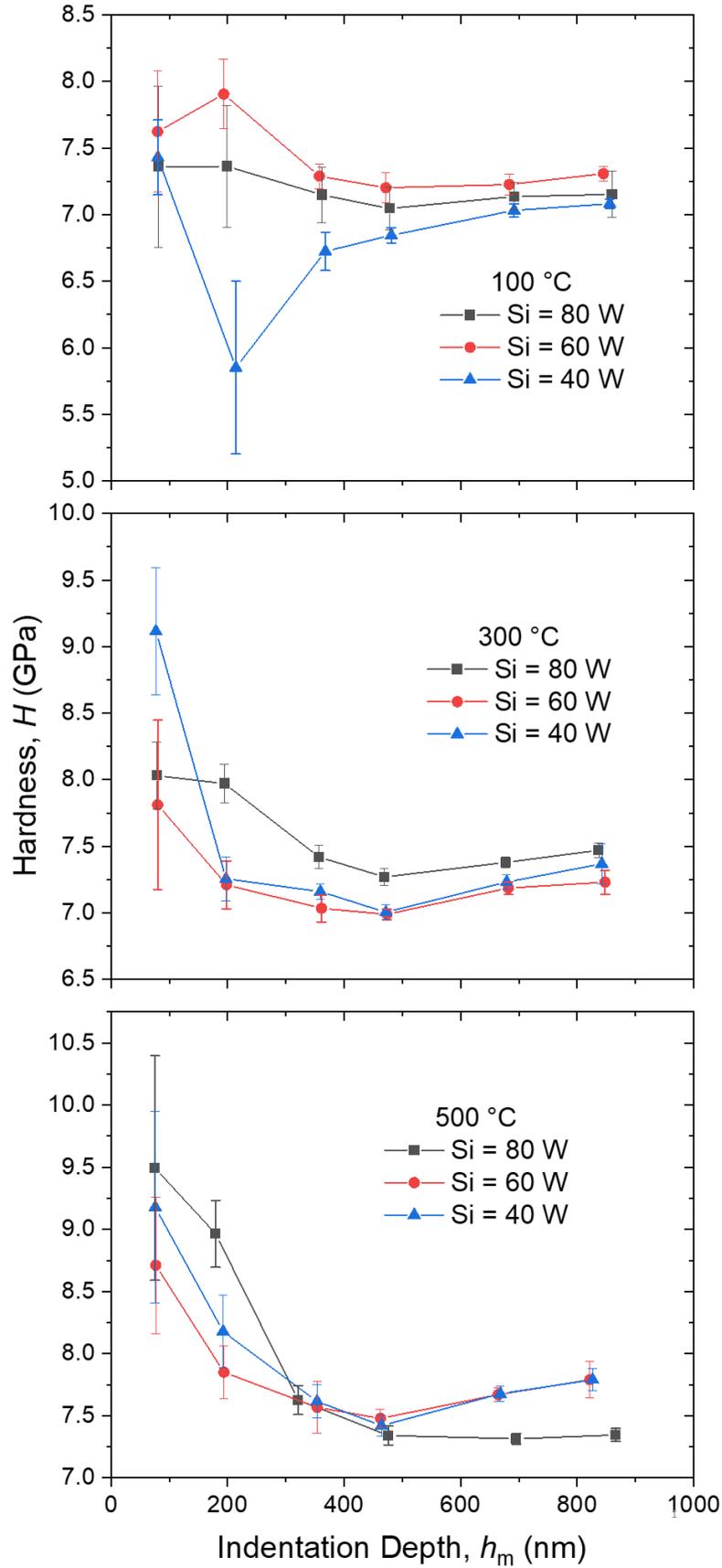

Figure S1: Hardness ($H$) as a function of indentation depth ($h_m$) for different deposition temperatures and Si target power.

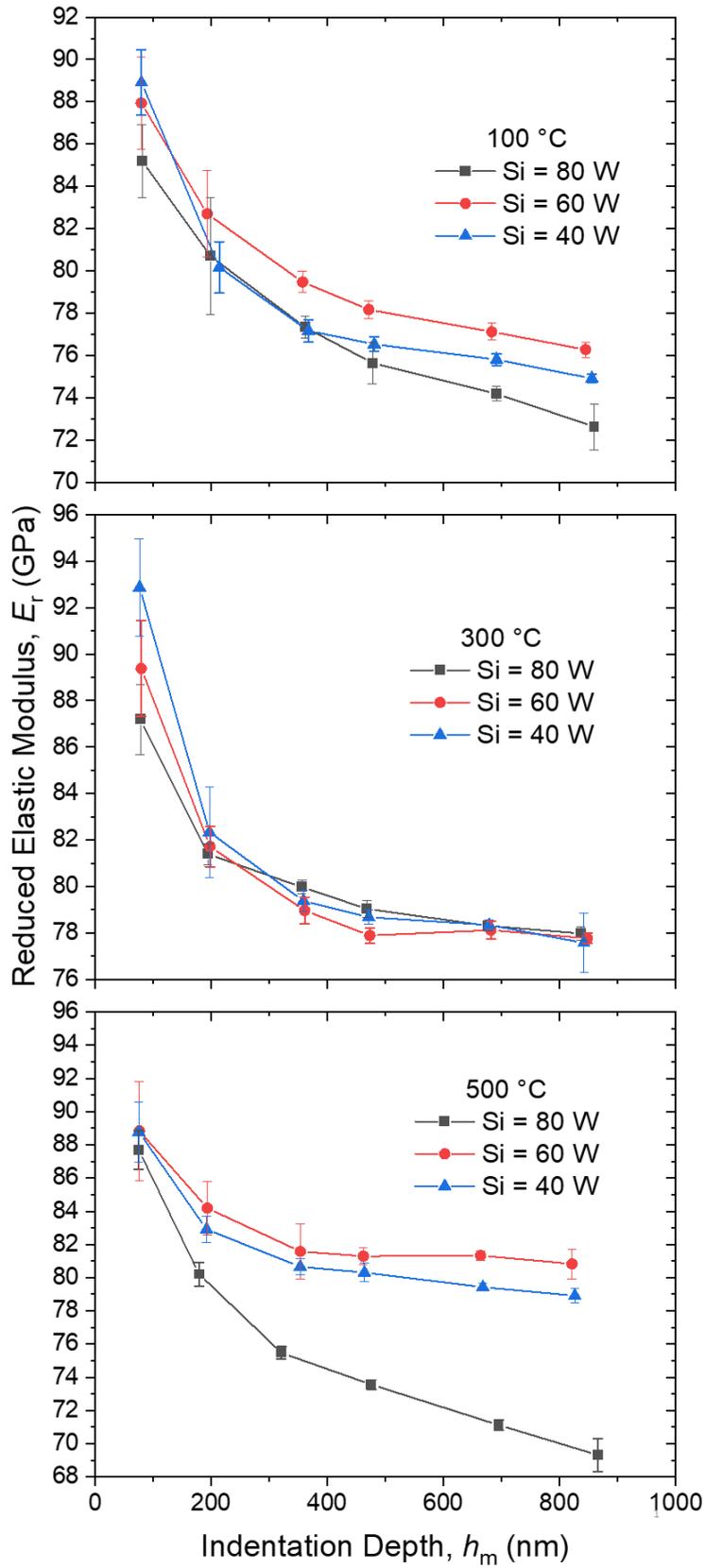

Figure S2: Reduced elastic modulus as a function of indentation depth for different deposition temperatures and Si target power.

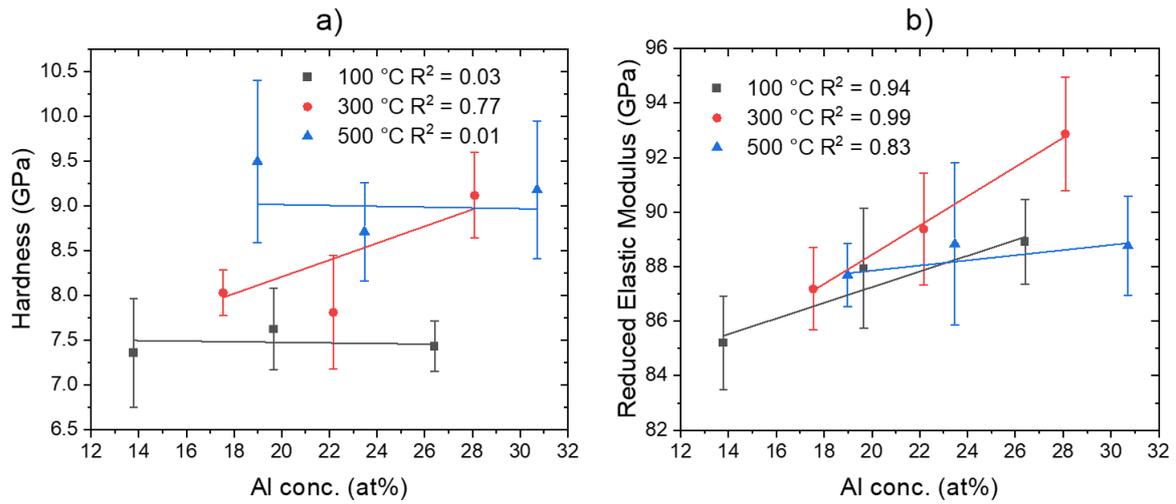

Figure S3: a) *H* and b) $E_r$ in substrate temperature series as a function of the Al content in at%.

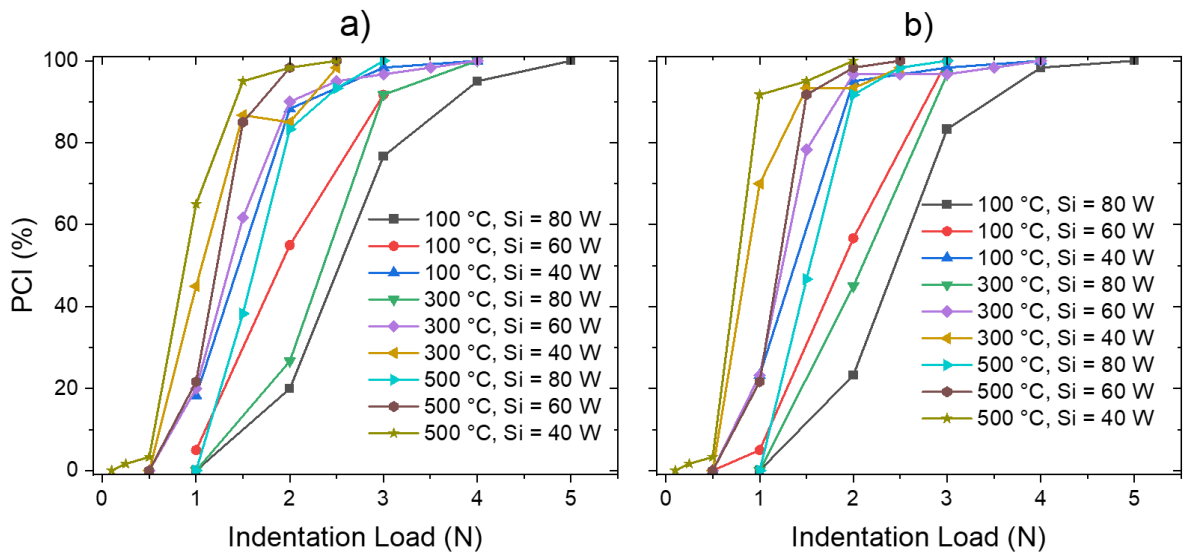

Figure S4: Probability of crack initiation (PCI) as a function of load for a) immediate crack initiation (*i.e.*, directly after indentation measurements) and b) delayed crack initiation (after >12 hours).

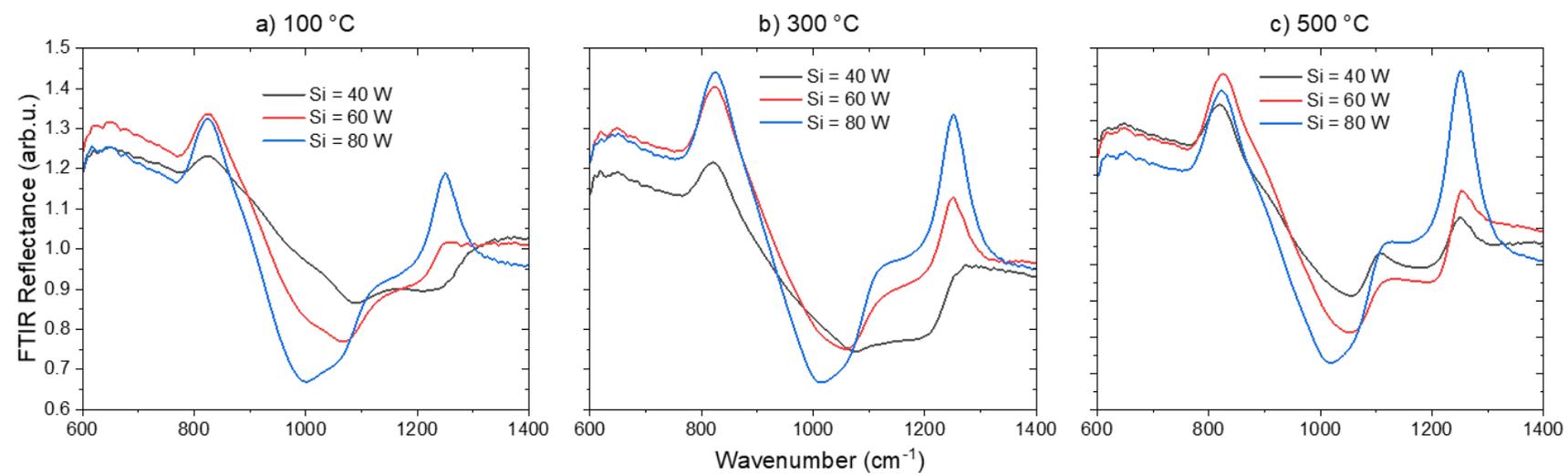

Figure S5: FTIR Reflectance spectra as measured (no baseline correction).

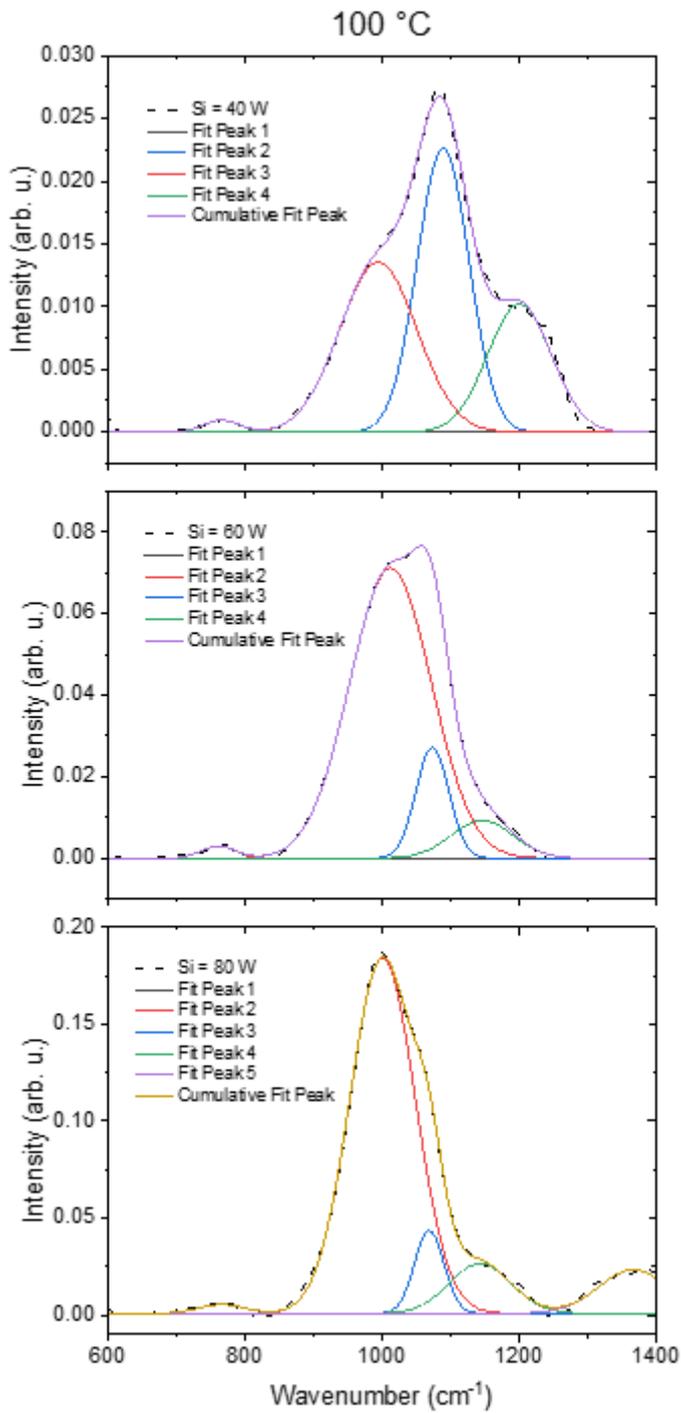

Figure S6: Deconvolution of the baseline-corrected KM transformed spectra for the Al-Si-O thin films prepared with a substrate temperature of 100 °C.

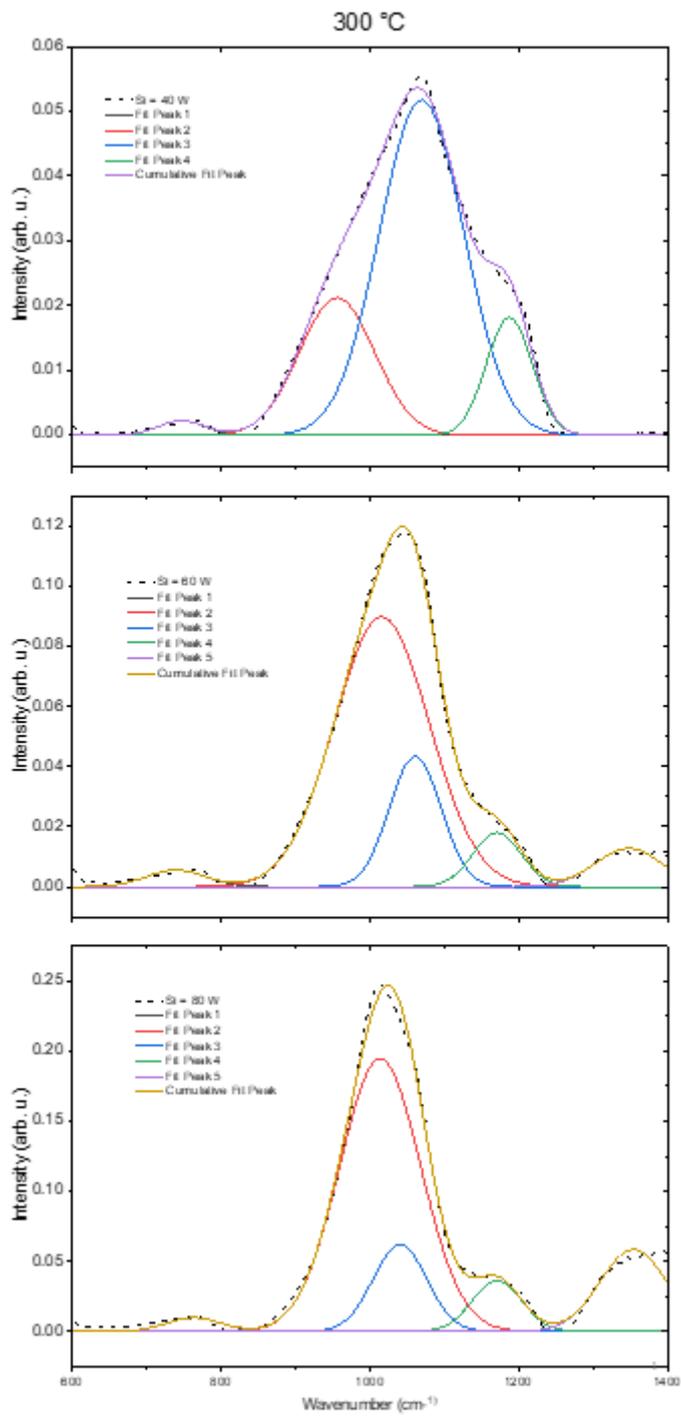

Figure S7: Deconvolution of the baseline-corrected KM transformed spectra for the Al-Si-O thin films prepared with a substrate temperature of 300 °C.

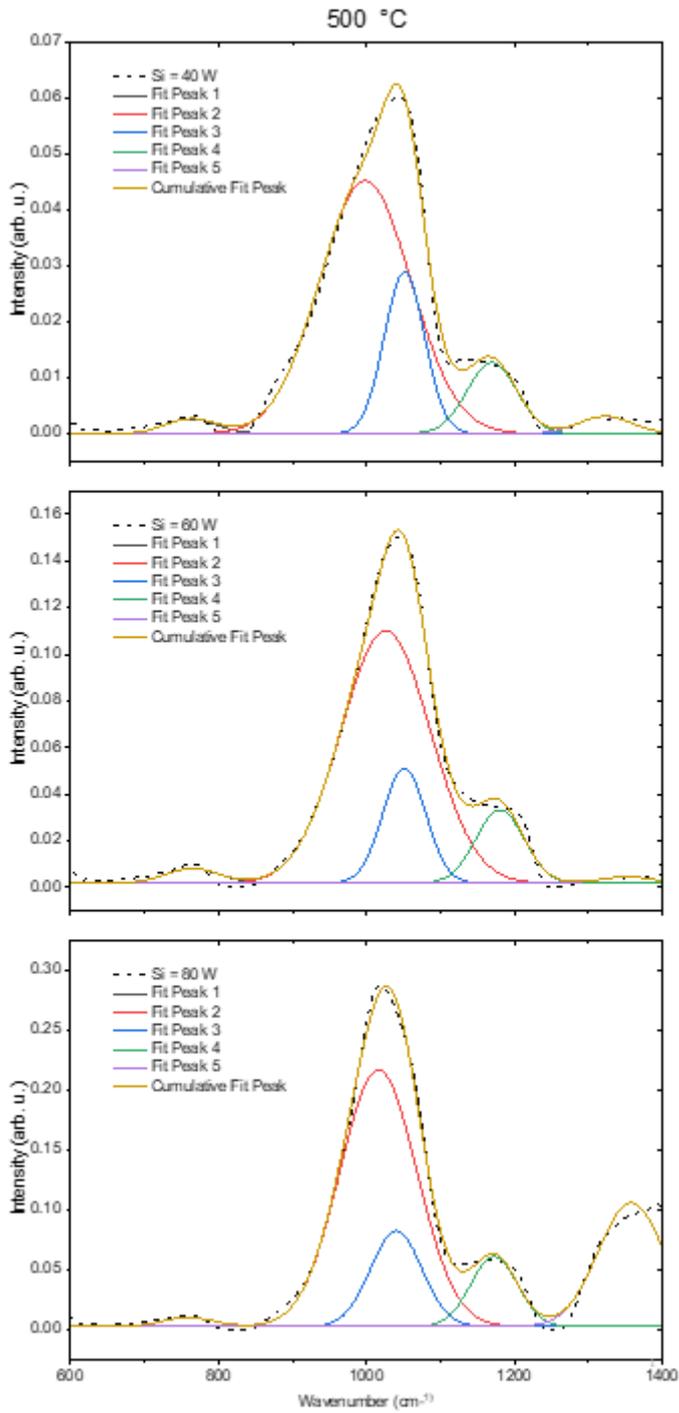

Figure S8: Deconvolution of the baseline-corrected KM transformed spectra for the Al-Si-O thin films prepared with a substrate temperature of 500 °C.